\documentclass[a4paper]{PoS}
\pdfoutput=1

\usepackage{amssymb, fontenc, times, mathptmx, graphicx}

\usepackage{amsmath}

\newcommand{\be}{\begin{equation}}
\newcommand{\ee}{\end{equation}}
\newcommand{\bea}{\begin{eqnarray}}
\newcommand{\eea}{\end{eqnarray}}

\title{Robinson-Trautman spacetimes and gauge/gravity duality}

\ShortTitle{Robinson-Trautman spacetimes and gauge/gravity duality}

\author{\speaker{Kostas Skenderis}\\
       Mathematical Sciences and STAG Research Centre, University of Southampton, Highfield, Southampton SO17 1BJ, United Kingdom\\
       E-mail: \email{K.Skenderis@soton.ac.uk}}

\author{Benjamin Withers\\
        Department of Theoretical Physics, University of Geneva,
24 quai Ernest-Ansermet, 1211 Gen\`eve 4, Switzerland\\
        E-mail: \email{benjamin.withers@unige.ch}}

\abstract{
We study far-from-equilibrium field theory dynamics using gauge/gravity duality applied to the Robinson-Trautman (RT) class of spacetimes and
we present a number of new results.  First, we assess the applicability of the hydrodynamic approximation to inhomogeneous plasma dynamics dual to RT spacetimes.
We prove that to any order in a late time expansion it is possible to identify variables corresponding to the local energy density and fluid velocity. However, we show using numerical examples that this does not hold at the non-perturbative level; for sufficiently inhomogeneous initial data a local rest frame does not exist.  
Second, we preset a new class of holographic inhomogeneous plasma flows on the plane. The corresponding spacetimes are not of the RT type but they can be obtained from RT spacetimes with spatially compact boundaries by coordinate transformations which generate Poincar\'e patch-like coordinates with planar boundaries. We demonstrate the application of this procedure using numerical examples. }

\FullConference{Corfu Summer Institute 2016 "School and Workshops on Elementary Particle Physics and Gravity"\\
		31 August - 23 September, 2016\\
		Corfu, Greece}

\begin{document}

\section{Introduction}
The description of quantum systems far from equilibrium remains a challenging exercise.
The machinery of gauge/gravity duality has become an attractive tool in this arena, which is arguably at its most mature in addressing CFTs with gauge group $SU(N)$ where $N$ is large. In this context non-equilibrium dynamics in the CFT are encoded by the a theory of classical gravity in one higher dimension with a negative cosmological constant, $\Lambda$. Often one also includes a non-extremal black hole which provides a thermal contribution.

There are by now many examples of the application of holography to non-equilibrium physics, with much attention being payed to sudden quenches of couplings or field theory sources, and the subsequent relaxation back to equilibrium (see for example \cite{Chesler:2008hg, Murata:2010dx, Balasubramanian:2011ur, Buchel:2013gba}). When $\Lambda<0$ there is no null infinity to which radiation can escape, and so -- in the absence of external dissipative couplings -- the spacetime settles down entirely due to the dissipation provided by the black hole horizon. These late time dynamics are in part described by quasi-normal modes of the black hole. At sufficiently late times an effective theory of the slowest ($\omega, k \ll T$) modes of the black hole persists, which takes the simple form of relativistic hydrodynamics, and equilibrium is ultimately reached. Other examples involve the construction of states which never return to equilibrium, such as the recent interest in the Riemann problem for CFTs considered both hydrodynamics and holography \cite{Bernard:2012je,2013JPhA...46K2001B,Chang:2013gba,Bhaseen:2013ypa,Amado:2015uza,Bernard:2015bba}, and the construction of stationary inhomogeneous plasma flows such as \cite{Figueras:2012rb}. In such cases the chief interest is the existence of steady-state regions which are also out of equilibrium.

Whilst the use of gauge/gravity duality provides a considerable computational advantage over a direct field theory computation, one is still faced with solving the Einstein equations. For generic initial data one may utilise the tools of numerical relativity. In these proceedings we consider instead a specialised class of metrics: the Robinson-Trautman (RT) spacetimes \cite{RT}. These are solutions to the Einstein equations for any $\Lambda$, but we focus exclusively on the case $\Lambda < 0$. They are inhomogeneous and time dependent, and are governed by a volume-preserving geometric flow equation for the spatial metric at the boundary of spacetime, the Calabi flow. This geometric flow is a fourth-order nonlinear diffusion process in $2+1$-dimensions, and its relative simplicity allows for valuable analytic control in the absence of any bulk isometries.

\emph{\textbf{Summary of previous results.}} The study of the holographic aspects of RT was initiated in \cite{deFreitas:2014lia, Bakas:2014kfa}. 
The RT metrics can be viewed as the nonlinear extension of algebraically special perturbations of Schwarzschild black holes. 
The algebraically special modes are purely dissipative, and when $\Lambda = 0$ describe radiation escaping to null infinity. When $\Lambda < 0$ the modes remain purely outgoing, and describe radiation which deforms the conformal boundary metric. As such, the Calabi flow equation determines the evolution of both the boundary metric and the interior of the spacetime, including expectation values of the stress tensor -- explicit expressions for these quantities in terms of the Calabi-flow variables are given in \cite{deFreitas:2014lia, Bakas:2014kfa}. 

The dissipation in RT is due to the outgoing radiation rather than absorption by the black hole. An interpretation of this is the dissipation that arises due to external couplings. As such the physics is somewhat different to the usual studies of out of equilibrium physics in holography. It is therefore important to analyse how RT behaves at late times, since its late time behaviour will no longer be governed by quasi-normal modes of a black hole. In cases where the boundary metric is topologically $\mathbb{R}\times S^2$, an analysis was made in \cite{Bakas:2014kfa}, finding exponentially damped modes with the Schwarzschild solution reached at late times. This has its origins in the late time properties of the Calabi-flow on $S^2$ \cite{Chrusciel:1992tj, Schmidt, Rendall, Singleton, Chrusciel:1992cj, Chrusciel:1992rv,Chen,Struwe}.

For non-compact boundaries, a study was initiated in \cite{Bakas:2015hdc} for the case $\mathbb{R}^{1,2}$. In general, at late times Schwarzschild is not reached, and instead the late time behaviour is governed by families of self-similar attractor solutions. For instance, depending on the boundary conditions chosen at infinity, a class of solutions may be approached that depends only on $\mu = x/t^{1/4}$. These solutions relate to the Riemann problem in the context of the field theory duals to RT spacetimes. The late time attractor solutions are governed by an underlying $z=4$ Lifshitz symmetry (which is subsequently broken at finite $T$), and the approach of a general dynamical evolution towards them is power-law in time, with the powers determined by the spectrum of operators about this Lifshitz invariant fixed point.

\emph{\textbf{Summary of new results in these proceedings.}} Sufficiently close to equilibrium, the theory of relativistic conformal hydrodynamics should provide a good approximation of the dynamics of the field theory. In particular, the expectation is that hydrodynamics should apply when all gradients are small compared with the temperature scale. At least, this regime will occur at sufficiently late times in unforced situations described through holography via the Einstein equations, since it is expected that a final stationary black hole will be reached, corresponding to equilibration. As discussed, this is the case for the RT spacetimes in the $S^2$ case, which approach Schwarzschild at late times.

However, for a fully nonlinear evolution of the Einstein equations, it is useful to assess the validity of the hydrodynamic approximation. Indeed there have been some surprises, with hydrodynamics applicable in regimes where it would naively be expected not to apply (for a recent example see \cite{Attems:2017ezz}). On the other hand, nonlinear dynamics may also produce regions with no local rest frame and local energy density \cite{Arnold:2014jva}.
The RT spacetimes provide a convenient way to test these issues, given the relative ease of obtaining a genuinely nonlinear inhomogeneous time dependent bulk solution. We will discuss this in section \ref{hydrosec}.

One goal of out of equilibrium holography is to understand the dynamics of a plasma on the plane. In order to study plasma dynamics on the plane, one needs an evolving bulk spacetime where the conformal boundary is planar. One efficient way to construct such solutions is to take known solutions in global-AdS coordinates where the boundary is the Einstein static universe, and perform a coordinate transformation such that the boundary is Minkowski space. Such a chart covers only part of the original spacetime, and corresponds to a conformal transformation from the boundary point of view. For instance the case of Schwarzschild-AdS perturbed by quasi-normal modes was considered in \cite{Friess:2006kw}.

We would like nontrivial plasma dynamics encoded by the RT solutions with boundary $\mathbb{R}^{1,2}$. One way to get this is to consider Calabi-flow on $\mathbb{R}^2$ as considered in \cite{Bakas:2015hdc}. Another approach is as described above, to consider an RT solution with conformal boundary of topology $\mathbb{R}\times S^{2}$ supplemented by a coordinate transformation which gives a bulk solution boundary with topology $\mathbb{R}^{1,2}$. The new coordinates can be thought of as Poincar\'e patch-like coordinates, but note that the new boundary metric of topology $\mathbb{R}^{1,2}$ is deformed by the dynamics associated to the outgoing radiation, just as in the $\mathbb{R}\times S^{2}$ case. Solutions so obtained describe nonlinear plasma dynamics on the plane which do not themselves correspond to a planar RT evolution. The details of the transformations and some numerical examples are given in section \ref{poincaresec}.

We begin with a brief review of the RT solution in section \ref{recap}. 

\begin{quote}
\emph{Note.} The work described here is part of a bigger program that was initiated by Ioannis Bakas, aiming to explore the connection between the Robinson-Trautmann spacetimes, Calabi flow and holography. Unfortunately, Ioannis passed away before we could complete this program. Ioannis was an outstanding physicist and a very good friend and he will be missed by the community. We would like to acknowledge his many contributions and we hope to be able to finish 
the work we started together. 
\end{quote}

\section{Brief review of the RT metrics and Calabi-flow\label{recap}}
The RT metric can be written, 
\bea
ds^2_{RT} &=& -F du^2 - 2 du dr + r^2 g_{ab} dx^a dx^b,\label{RT1}\\
g_{ab}dx^a dx^b &=& \frac{1}{\sigma^2}d\Sigma_k^2,\label{RT2}\\
F &\equiv & -\frac{\Lambda}{3}r^2  -2 r\frac{\partial_u \sigma}{\sigma}+ \frac{R_g}{2}- \frac{2m}{r}.\label{RT3}
\eea
$R_g$ is the Ricci scalar of the metric $g$,  $\sigma$ is a function of $u$ and the two $x^a$, $d\Sigma_k^2$ denotes the line element for an Euclidean 2-metric with constant scalar curvature $2k$ so that $k=0$ is the plane or torus, $k=1$ is the unit round sphere and $k=-1$ hyperbolic space. Additionally the solution has a constant mass parameter $m$. The RT metric \eqref{RT1}-\eqref{RT3} solves the Einstein equations with cosmological constant $\Lambda$ provided $g$ obeys the Calabi-flow equation, 
\be
\partial_u g_{ab} = \frac{1}{12m} \nabla^2_g R_g\, g_{ab}, \label{CalabiK}
\ee
which, after inserting \eqref{RT2} can be seen to be a nonlinear fourth-order diffusion equation for $\sigma$. Here $\nabla_g^2$ and $R_g$ are the Laplacian and Ricci scalar for $g$.

In the case $\Lambda < 0$ the conformal boundary metric -- reached as $r\to\infty$, where the outgoing coordinate $u$ becomes boundary time coordinate $t$ -- is given by
\be
ds^2 = -dt^2 + \frac{L^2}{\sigma^2}d\Sigma_k^2,
\ee
where $L^2 = -3/\Lambda$. Explicit expressions for the expectation value of the dual field theory stress tensor can be found in \cite{Bakas:2014kfa} depending on up to six spatial derivatives of $\sigma$.

\section{Validity of the hydrodynamic approximation\label{hydrosec}}
In this section we assess the validity of the hydrodynamic approximation applied to evolutions described by RT spacetimes. In particular, in order for a hydrodynamic description to hold, we must first identify the local fluid variables: an energy density $\epsilon(t,x)$ and a 4-velocity field $u^\mu(t,x)$. To do so we look for solutions of the eigenvalue problem, 
\be
T^{\mu}_{\phantom{\mu}\nu} u^\nu = - \epsilon u^\mu\qquad \text{with}\quad u^\mu u_\mu = -1 \label{eigenvalueproblem}
\ee
where $T_{\mu\nu}$ is the expectation value of the field theory stress tensor, which is provided here by holography from the near boundary behaviour of RT solutions. In domains where \eqref{eigenvalueproblem} yields physical solutions for $\epsilon, u^\mu$, we may compare with the expected result from conformal relativistic hydrodynamics, whose constitutive relations are
\be
(T_{\text{hydro.}})^\mu_{\phantom{\mu}\nu} = \frac{3}{2}\epsilon u^\mu u_\nu + \frac{1}{2}\epsilon \delta^\mu_\nu + \Pi^\mu_{\phantom{\mu}\nu}, \label{thydro}
\ee
where $\Pi^\mu_{\phantom{\mu}\nu}$ represents higher order derivative corrections to the ideal hydrodynamic behaviour, and where we work in the Landau frame $\Pi^\mu_{\phantom{\mu}\nu}u^\nu = 0$. Note also that $\Pi^\mu_{\phantom{\mu}\mu} = 0$, which is due  to the conformal invariance of the dual QFT.

In section \ref{latetimeeval} we prove that it is possible to obtain local hydrodynamic variables to all orders in a perturbative late time expansion. However, interestingly this does not continue to hold at the non-perturbative level, which we demonstrate with explicit numerical solutions in section \ref{evalnumerics}. Similar results were reported in \cite{Arnold:2014jva}.
We also use the numerical solutions to demonstrate the expected convergence to hydrodynamics at late times as governed by the late time expansion.

For simplicity we focus the Calabi flow on $S^2$ with axial symmetry. Thus we are constructing RT spacetimes with a conformal boundary of the form $\mathbb{R}\times S^2$. The solutions will be functions of boundary time $t$ and the polar angle $\theta$.

\subsection{Proof of solution to eigenvalue problem \eqref{eigenvalueproblem} to any order in a late time expansion\label{latetimeeval}}

Details of the late time expansion for the axially symmetric $S^2$ case were given in \cite{Bakas:2014kfa}. Here we utilise these results to prove that the eigenvalue problem \eqref{eigenvalueproblem} can be solved in a late time expansion to any order. As in \cite{Bakas:2014kfa} we adopt a normalisation where $\lim_{t\to\infty} \sigma  = 1$ in this section. 

First we note that for the late time expansion of $\sigma$, 
\be
\sigma(x,u) = 1 + \sum_{n} \sigma^{(n)}
\ee
we can obtain the boundary metric $g_{ab}$ and holographic stress tensor $T_{ab}$ in that expansion. The index in parentheses indicates the order in the late time expansion, i.e. $\sigma^{(n)} \sim e^{-2n\frac{u}{m}}$, and explicit expressions for the first few coefficients can be found in  \cite{Bakas:2014kfa}.

Our goal here is to show that we can always find hydrodynamic variables to solve
\be
T_{ab} = T^{\text{hydro.}}_{ab} \label{TT}
\ee
order by order, thereby solving \eqref{eigenvalueproblem}. We begin by expanding the hydrodynamic variables in a late time expansion, first the velocity,
\be
u_a = u_a^{(0)} + u_a^{(1)} + u_a^{(2)} + \ldots
\ee
and similarly for the energy density, $\epsilon$. Similarly, we expand the dissipative contribution to the stress tensor in this way, 
\be
\Pi_{ab} = \Pi^{(0)}_{ab} + \Pi^{(1)}_{ab} + \Pi^{(2)}_{ab} + \Pi^{(3)}_{ab} + \ldots.
\ee

We begin by considering explicitly the solution to a few orders in the expansion. By explicit computation we have  \cite{Bakas:2014kfa}
\be
u^{(0)}_t = -1, \quad u^{(0)}_\theta = 0, \quad u^{(0)}_\phi = 0, \quad \Pi^{(0)}_{ab} = 0,  \label{uzeros}
\ee
then $\epsilon^{(0)} = (u_a u_b T^{ab})^{(0)} = T^{tt}_{(0)}$. 

We now solve all first order quantities.
Knowing \eqref{uzeros}, allows us to iteratively construct $\Pi_{(n)}^{tb}$ given the Landau frame condition, $u_a \Pi^{ab} = 0$. At first order we have
\be
(u_a \Pi^{ab})^{(1)} = u^{(0)}_a \Pi^{ab}_{(1)} = 0 \implies \Pi^{tb}_{(1)} =0.
\ee
Using this result and the fact that $T_{tt}^{(1)}$ vanishes \cite{Bakas:2014kfa} we then find that 
\be \label{e2}
\epsilon^{(1)} = \frac{3}{2} \epsilon^{(0)} u_t^{(1)}.
\ee
Then from the $t\theta$ and $t\phi$ components of \eqref{TT}  we obtain 
\be
u^{(1)}_\theta  = -\frac{2 T^{(1)}_{t\theta}}{3\epsilon^{(0)}}; \qquad u^{(1)}_\phi  = -\frac{2 T^{(1)}_{t\phi}}{3\epsilon^{(0)}}
\ee
We now use  $u_a u^a = -1$ to obtain
\be
u_t^{(1)}=0,
\ee
which upon use of (\ref{e2}) implies $\epsilon^{(1)}=0$.
The $\theta\theta$-component of \eqref{TT} yields
\be
\Pi_{\theta\theta}^{(1)} = T_{\theta\theta}^{(1)} - \frac{1}{2} T_{tt}^{(0)} g_{\theta\theta}^{(1)} 
\ee
and the  $\phi\phi$-component of \eqref{TT} yields
\be
\Pi_{\phi\phi}^{(1)} = T_{\phi\phi}^{(1)} - \frac{1}{2} T_{tt}^{(0)} g_{\phi\phi}^{(1)} 
\ee
So we managed to obtain all quantities, $u^{(1)}_a$, $\epsilon^{(1)}$, $\Pi^{(1)}_{ab}$, to this order.


The general strategy adopted to solve the eigenvalue problem to first order can be extended to give a proof to all orders, by induction. We have all quantities of interest at order 0, i.e. $u_a^{(0)}$, $\epsilon^{(0)}$ and $\Pi_{ab}^{(0)}$. Next suppose that we have determined everything at order $(n-1)$ and below, i.e. $u^{(n-1)}_a$, $\epsilon^{(n-1)}$, $\Pi^{(n-1)}_{ab}$ and lower order terms. The following steps can be used to advance to order $n$:  
\begin{enumerate}
\item Solve the Landau frame condition $u_a \Pi^{ab} = 0$ for $\Pi^{(n)}_{ta}$. 
\item From the $tt$-component of \eqref{TT} solve for $\epsilon^{(n)}$ in terms of the unknown $u_t^{(n)}$.
\item From the $t\theta$-component of \eqref{TT} solve for $u_\theta^{(n)}$ in terms of the unknown $u_t^{(n)}$.
\item From the $t\phi$-component of \eqref{TT} solve for $u_\phi^{(n)}$ in terms of the unknown $u_t^{(n)}$.
\item From the condition $u_a u^a = -1$ solve for $u_t^{(n)}$.
\item From the $\theta\theta$-component of \eqref{TT} solve for $\Pi_{\theta\theta}^{(n)}$. 
\item From the $\phi\phi$-component of \eqref{TT} solve for $\Pi_{\phi\phi}^{(n)}$. 
\end{enumerate}
With the above results we now have all quantities at order $n$, i.e. $u_a^{(n)}$,  $\epsilon^{(n)}$ and $\Pi_{ab}^{(n)}$. This completes the proof by induction.

Since this is a perturbative construction, $\epsilon$ will stay positive within perturbation theory, and we would need to resum the series in order to see any possibility of regions where $\epsilon < 0$.

\subsection{Numerical examples\label{evalnumerics}}
In this section we use numerical solutions of the Calabi-flow to assess the hydrodynamic approximation, both for the existence of solutions to the \eqref{eigenvalueproblem} away from the late time regime (considered separately in section \ref{latetimeeval}), and also to illustrate the approach to the late time regime itself. Note that existence of solutions to the \eqref{eigenvalueproblem} can be immediately assessed without performing any evolution, simply by looking at initial data at $t=0$, which is sufficient to obtain $T_{\mu\nu}(0,x)$, where we write $x=\cos\theta$ for convenience. We pick a class of initial data which characterise deviations from equilibrium using axially-symmetric ($m=0$) spherical harmonics, 
\be
\sigma(0,x) = 1 + c Y^0_\ell(x) \label{initialdata}
\ee
where $c$ is a constant. We have omitted the redundant azimuthal angle argument. We also restrict attention to $\ell$-even in order to have an antipodal symmetry. For $\ell = 2$ we find that there is a critical value $c^\ast_2 \simeq 0.32$, such that for $0<c<c^\ast_2$ all spatial points admit real solutions to \eqref{eigenvalueproblem} whilst for $c>c^\ast_2$ there are points which do not. Similarly for the case $\ell = 4$ we find a critical value $c^\ast_4 \simeq 0.020$. 

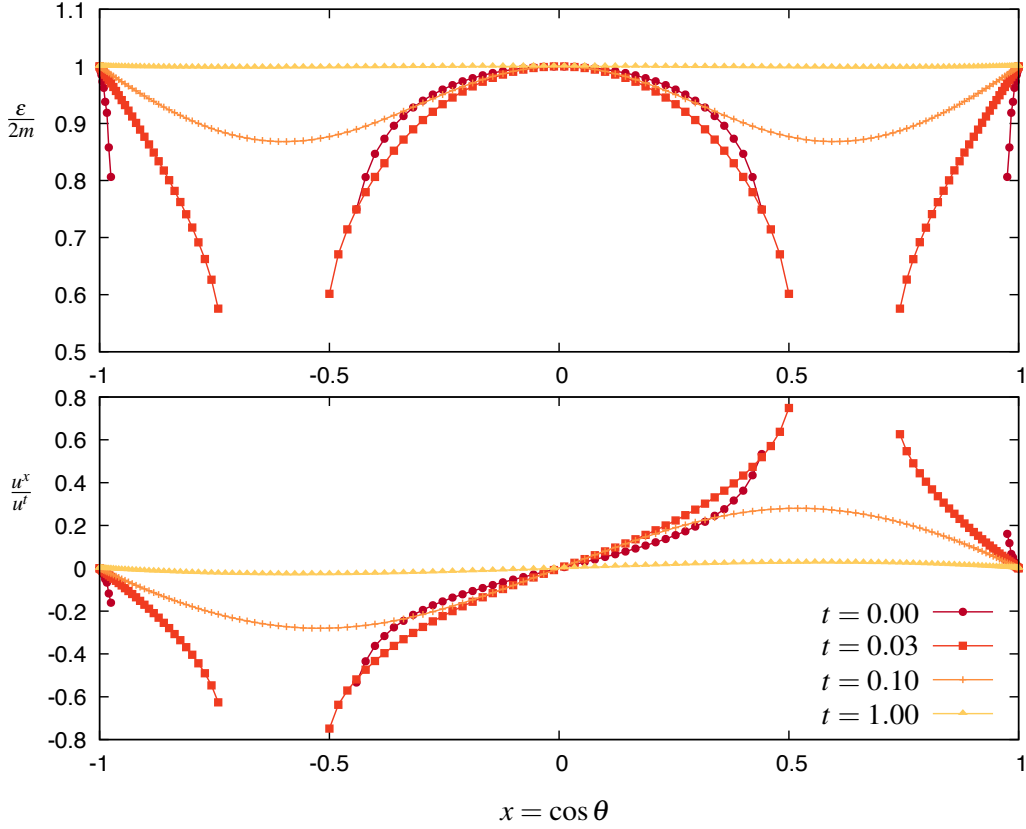
\begin{figure}[h!]
\begin{center}
\includegraphics[width=0.9\textwidth]{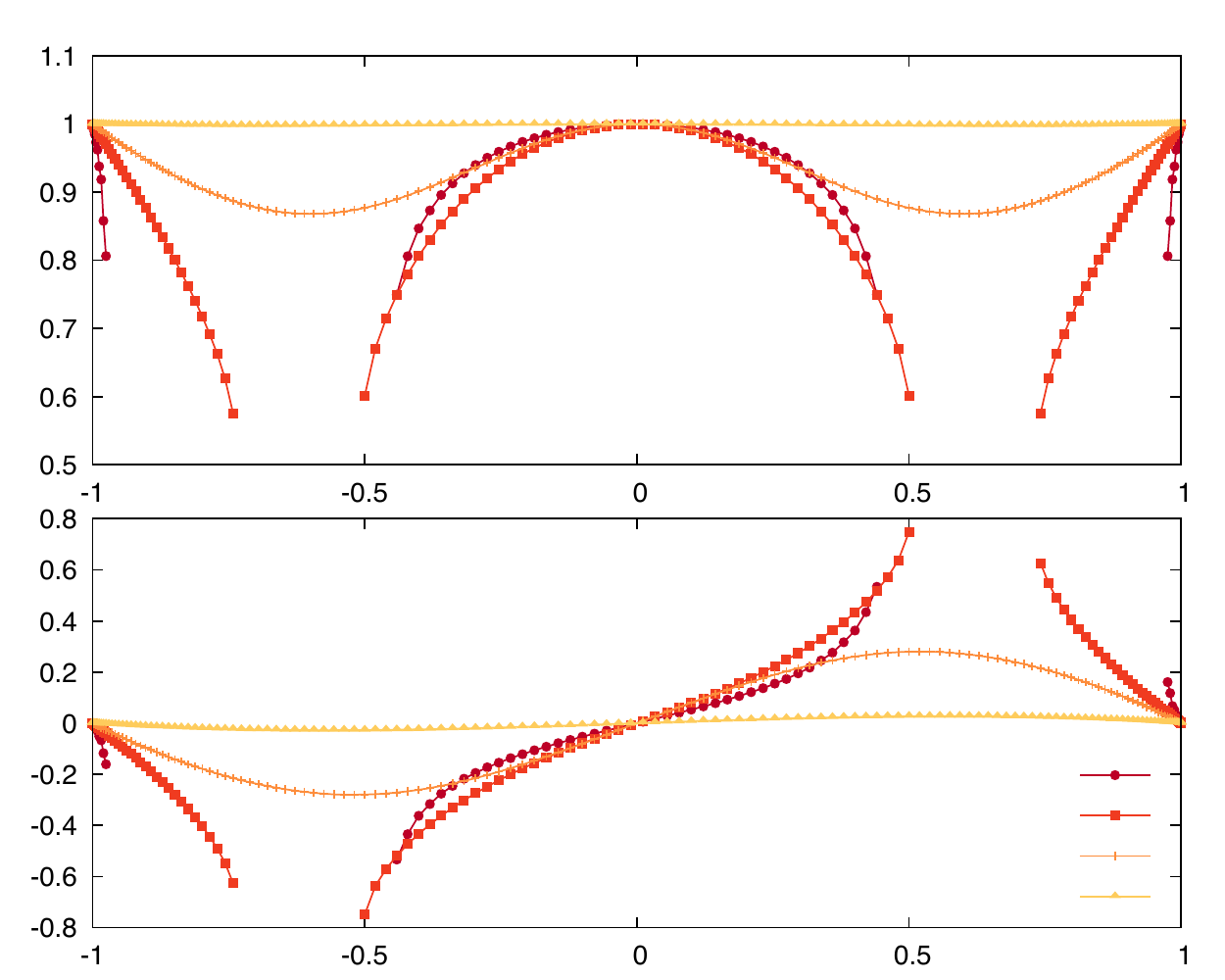}
\begin{picture}(0.1,0.1)(0,0)
\put(-190,-10){\makebox(0,0){$x = \cos\theta$}}
\put(-390,110){\makebox(0,0){$\frac{u^x}{u^t}$}}
\put(-390,250){\makebox(0,0){$\frac{\epsilon}{2m}$}}
\put(-72, 64){\makebox(0,0){$t=0.00$}}
\put(-72, 52){\makebox(0,0){$t=0.03$}}
\put(-72, 39){\makebox(0,0){$t=0.10$}}
\put(-72, 26){\makebox(0,0){$t=1.00$}}
\end{picture}
\vspace{1em}
\caption{The local fluid energy density (upper panel) and velocity (lower panel) corresponding to a RT spacetime with initial data \eqref{initialdata} at $\ell =2$ and $c=1/2$. The points correspond to fixed $t$ slices, with $t=0$ (dark red circles), $t=0.03$ (orange squares), $t=0.10$ (light orange crosses), $t=1.00$ (yellow triangles). For the two earliest slices shown there are two belts on the sphere where no real solution to the eigenvalue problem (\ref{eigenvalueproblem}) exist. These regions narrow and disappear, approaching equilibrium at late times where they are everywhere defined.\label{slices} }
\end{center}
\end{figure}

Let us now turn to the time evolution for the case $c=1/2 > c^\ast_2$ and $\ell = 2$. The angular profile of $\epsilon$, where defined, is displayed in figure \ref{slices} for a handful of different time slices. Initially there are two `belts' on the sphere where there is no real solution to \eqref{eigenvalueproblem}. The fluid flows towards the poles in each hemisphere, with zero velocity on the equator and at the pole itself.

We would like to quantify the approach to equilibrium at late times in terms of the rest frame variables we have identified. To this end, once we have $\epsilon, u^\mu$, we can construct two 3-vectors $n_I$ where $I=1,2$ such that,
\be
n_I \cdot n_J = \delta_{IJ} \qquad u\cdot n_I = 0,
\ee
and use them to define two pressures,
\be
p_1 = n_1\cdot T\cdot n_1\qquad p_2 = n_2\cdot T\cdot n_2.
\ee
Additionally we may construct a strain $n_1\cdot T\cdot n_2$ but this vanishes for the axially symmetric case of interest here. Here we take $n_2 \propto \partial_\phi$. We can identify $\epsilon, p_1, p_2, u^\mu$ from the numerical solutions, and we can obtain the same quantities analytically in a late time expansion by following the methodology outlined in section \ref{latetimeeval}. This allows us to compare the numerical evolution of rest frame variables with the late time expansion. In the case of antipodal symmetry, the first 14 orders in the late time expansion are determined by a single mode, the quadrupole ($\ell = 2$), which at linear order takes the form, 
\be
\delta\sigma(u,x) =  a\left(x^2-\frac{1}{3}\right) e^{-\frac{2\sigma_\infty^4}{m} u},
\ee
where $\sigma_\infty$ is the constant late time value of $\sigma$.
New data, $b$, can enter at order $15$ corresponding to the contribution of a linear hexadecapole ($\ell = 4$) mode,
\be
\delta\sigma(u,x) =  b\left(\frac{35}{8}x^4-\frac{15}{4}x^2+\frac{3}{8}\right) e^{-\frac{30\sigma_\infty^4}{m} u},
\ee
which goes in hand with a term $\sim a^{15} \left(\frac{35}{8}x^4-\frac{15}{4}x^2+\frac{3}{8}\right) u\, e^{-\frac{30\sigma_\infty^4}{m} u}$ due to the overlap of the two mode expansions  \cite{Bakas:2014kfa}.

\begin{figure}[h!]
\begin{center}
\includegraphics[width=0.8\textwidth]{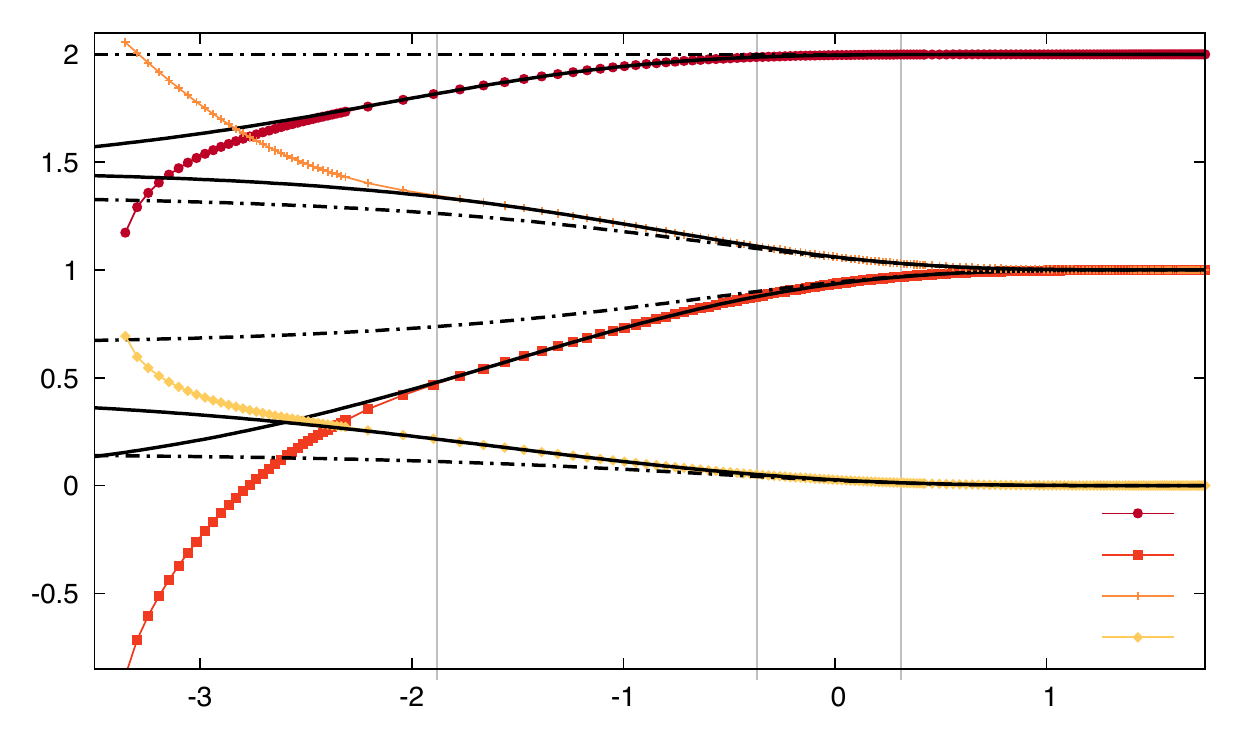}
\begin{picture}(0.1,0.1)(0,0)
\put(-170, -5){\makebox(0,0){$\log t$}}
\put(-360,146){\makebox(0,0){$\frac{\epsilon}{m}, \frac{p_1}{m}, \frac{p_2}{m}, \frac{u^x}{u^t}$}}
\put(-58,65){\makebox(0,0){$\epsilon/m$}}
\put(-60,54){\makebox(0,0){$p_1/m$}}
\put(-60,43){\makebox(0,0){$p_2/m$}}
\put(-60,31){\makebox(0,0){$u^x/u^t$}}
\put(-214, 16){\makebox(0,0){\scriptsize{$\log t_{energy}$}}}
\put(-135, 16){\makebox(0,0){\scriptsize{$\log t_{hyd}$}}}
\put(-95, 16){\makebox(0,0){\scriptsize{$\log t_{iso}$}}}
\end{picture}
\vspace{1em}
\caption{Approach to equilibrium described by local fluid variables $\epsilon, u^\mu$ and associated pressures $p_1$ and $p_2$. The points correspond to numerical solutions, for which $\epsilon, u^\mu$ are obtained by solving the eigenvalue problem \eqref{eigenvalueproblem}. The black dashed and black solid lines are the same quantities computed order-by-order in a late time expansion. The black dashed lines give the linear quadrupole contribution, whilst the solid black line gives its nonlinear completion to order 15 in the late time expansion. The initial data is as given in \eqref{initialdata} with $\ell = 2, c = 1/2$ and we here show the the quantities at $x\simeq 0.61$ -- similar agreement is seen at other locations. For earlier times not shown there is no global solution to the eigenvalue problem \eqref{eigenvalueproblem}. 
\label{eppplot}}
\end{center}
\end{figure}
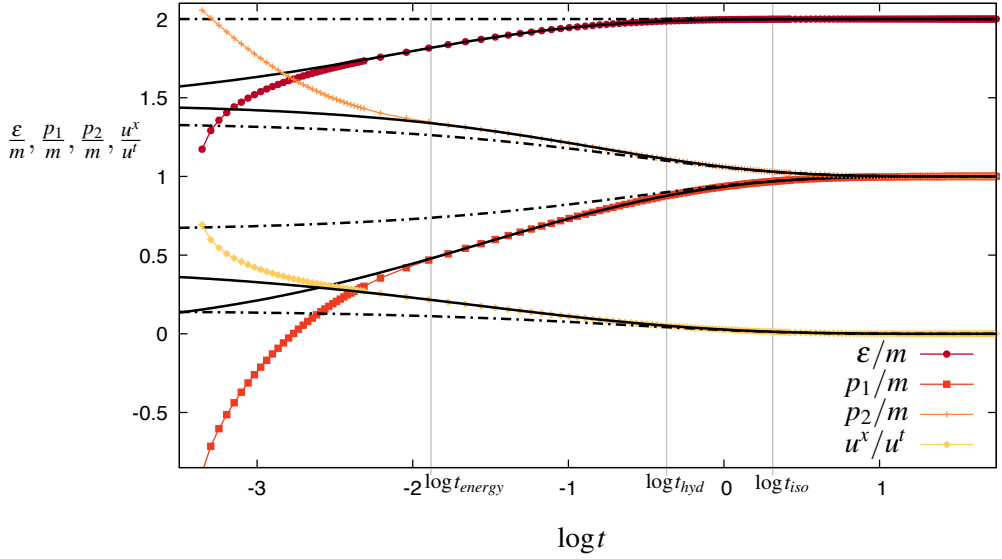

In figure \ref{eppplot} we show the numerical evolution of $\epsilon, p_1, p_2, u^x/u^t$ and a comparison to the linear quadrupole mode, as well as its nonlinear extension up to order 15 in the late time expansion. For this particular numerical solution we find $\sigma_\infty \simeq 0.97$ and the late time values of the fluid quantities are as expected, i.e. $\epsilon = 2m$, $p_1=m$, $p_2=m$, $u^x/u^t =0$. Late time deviations are governed by the amplitude of the linear quadrupole mode, which we find to be $a \simeq 0.32$. Beyond order $\sim8$ in the late time expansion the agreement with the fluid variables is not visibly improved by adding higher order terms in this case, and we have not been able to reliably extract the hexadecapole amplitude $b$ at order 15 from our numerics.

We now make some observations on the timescales indicated by the rest frame variables displayed in figure \ref{eppplot}. The first observation is that the first order late time expansion (the dashed lines), which describes a hydrodynamic regime \cite{Bakas:2014kfa}, provides a good description of the evolution earlier than the time at which the pressures become equal. Such behaviour was observed in other conformal systems \cite{Chesler:2010bi,Heller:2011ju}. To better characterise this behaviour we construct a `hydrodynamisation' time, $t_{hyd}$, here defined to be the time after which both pressures $p_1,p_2$ agree with the order-1 late time expression to better than 10\% for all $x$, i.e., we seek the time after which
\be
\max_x\left|\frac{p_I - p^{(1)}_I}{\bar{p}}\right| < \frac{1}{10}, \qquad\qquad (t_{hyd})
\ee
satisfied by both $I=1,2$ where $\bar{p} = (p_1+p_2)/2$ and the superscript in parentheses denotes the first-order expression in the late time expansion. For the numerical solutions shown in figures \ref{slices} and \ref{eppplot} we find $t_{hyd} \simeq 0.69$, indicated by a vertical grey line in figure \ref{eppplot}. 
The second observation is that the energy density reaches equilibrium earlier than the time at which the pressures become equal. To quantify this statement further, we introduce two timescales: an energy equilibration time, $t_{energy}$, and an isotropisation time $t_{iso}$. We define $t_{energy}$ to be the time after which $\epsilon$ differs from its eventual equilibrium value $\epsilon_{eq}$ (here $\epsilon_{eq} = 2m$) by less than 10\% for all $x$, i.e.
\be
\max_x\left|\frac{\epsilon - \epsilon_{eq}}{\epsilon}\right| < \frac{1}{10}. \qquad\qquad (t_{energy})
\ee
Note that for conformal systems, such as the one we are studying, $t_{energy}$ is the same as the timescale characterised by when $\bar{p}$ is within 10\% of its equilibrium value, $p_{eq}$ (here $p_{eq} = m$).
Next we define $t_{iso}$ to be the time after which the pressures differ from one another by less than 10\% for all $x$, i.e.
\be
\max_x\left|\frac{p_1-p_2}{\bar{p}}\right| < \frac{1}{10}. \qquad\qquad (t_{iso})
\ee
For the evolution shown in figures \ref{slices} and \ref{eppplot} we find $t_{energy} \simeq 0.15$ and $t_{iso} \simeq 1.37$, indicated by vertical grey lines in figure \ref{eppplot}. We note that $t_{energy} < t_{hyd} < t_{iso}$.

\section{Transformations to the Poincar\'e patch\label{poincaresec}}
In this section we seek to generate nontrivial plasma dynamics on $\mathbb{R}^{1,2}$, which requires a nontrivial evolving planar bulk spacetime. One way to achieve this is to consider directly RT metrics corresponding to Calabi flows on $\mathbb{R}^2$ as considered in \cite{Bakas:2015hdc}. Another approach which we adopt here, is to perform a coordinate transformation from RT solutions which have a boundary of topology $\mathbb{R}\times S^2$. The new coordinates correspond to a Poincar\'e patch-like slicing of the bulk spacetime. Transformations of this type were previously considered in their action on Schwarzschild perturbed by quasi-normal modes to generate inhomogeneous plasma flows on the plane \cite{Friess:2006kw}. 

\subsection{Generalities}
We begin by considering the near boundary metric in Fefferman-Graham form,
\be
ds^2 = L^2\left(\frac{dr^2}{r^2} + \frac{1}{r^2} \left(g_{(0)ab} + r^2 g_{(2)ab} + r^3 g_{(3)ab} + O(r^4)\right) dx^a dx^b \right)
\ee
where $L^2 = -3/\Lambda$. The stress tensor is given by \cite{deHaro:2000vlm},
\be
T_{ab} = -\frac{3}{2\kappa^2}L^2 g_{(3)ab}.
\ee
Expressions for the $g_{(i)ab}$ corresponding to RT solutions were computed in \cite{Bakas:2014kfa}. 

The Einstein universe $R \times S^2$ and $\mathbb{R}^{1,2}$ are conformally related and the conformal transformation can be implemented via a bulk diffeomorphism.
This diffeomorphism has been determined in the near boundary region in  \cite{Skenderis:2000in}.
Here we present a derivation following the approach set out in \cite{Friess:2006kw}.
Consider AdS$_4$ as an embedding in $\mathbb{R}^{2,3}$ defined by the hyperboloid,
\be
-X_{-1}^2 - X_0^2 + X_1^2 + X_2^2 + X_3^2  = -L^2.
\ee
The metric for global AdS in Fefferman-Graham coordinates can then be reached by
\bea
X_{-1} &=& L\left( \frac{1}{\rho}+\frac{\rho}{4}\right) \cos t\\
X_0 &=& L\left(\frac{1}{\rho}+\frac{\rho}{4}\right) \sin t\\
X_1 &=& L\left(\frac{1}{\rho}-\frac{\rho}{4}\right) \sqrt{1-\chi^2} \cos \phi\\
X_2 &=& L\left(\frac{1}{\rho}-\frac{\rho}{4}\right) \sqrt{1-\chi^2} \sin \phi\\
X_3 &=&L\left( \frac{1}{\rho}-\frac{\rho}{4}\right) \chi
\eea
giving 
\be
ds^2 = L^2\left( \frac{d\rho^2}{\rho^2}  +\frac{1}{\rho^2}\left(-\left(1+\frac{\rho^2}{4}\right)^2 dt^2 + \left(1-\frac{\rho^2}{4}\right)^2 d\Omega_2^2\right)\right)
\ee
where $d\Omega_2^2 = \frac{d\chi^2}{1-\chi^2} + (1-\chi^2)d\phi^2$ and where $\chi = \cos{\theta}$, so that, for the terms presented in the Fefferman-Graham expansion (with radial variable $r\to\rho$),
\bea
ds_{(0)}^2 &=& -dt^2 + d\Omega_2^2\\
ds_{(2)}^2 &=& -\frac{1}{2} dt^2 - \frac{1}{2} d\Omega_2^2\\
ds_{(3)}^2 &=& 0.
\eea
Note that the $t$ here differs from the one used in earlier sections by an overall normalisation by $L$.
Similarly, the metric for planar AdS with boundary polar coordinates $(R,\Phi)$ can be reached by
\bea
X_{-1} &=& L\frac{1 + R^2 + z^2 - \tau^2}{2 z}\\
X_{0} &=& L \frac{ \tau}{z} \\
X_{1} &=& L \frac{R}{z} \cos{\Phi}\\
X_{2} &=& L \frac{R}{z} \sin{\Phi}\\
X_{3} &=& L \frac{1-R^2-z^2+\tau^2}{2 z}
\eea
giving
\be
ds^2 = L^2\left(\frac{dz^2}{z^2} + \frac{1}{z^2}\left(-d\tau^2 + dR^2+R^2d\Phi^2\right)\right)
\ee
and again, for a Fefferman-Graham expansion  (with radial variable $r\to z$), we have
\bea
d\tilde{s}_{(0)}^2 &=& -d\tau^2 + dR^2+R^2d\Phi^2\\
d\tilde{s}_{(2)}^2 &=& 0\\
d\tilde{s}_{(3)}^2 &=& 0,
\eea
where the tilde notation has been introduced to indicate the planar case.

In order to perform the transformation of interest, we just need to find the conversion from $(t,\rho,\chi,\phi)$ to $(\tau,z,R,\Phi)$ by equating the $X$'s:
\bea
z &=& \frac{\rho}{\left(1-\frac{\rho^2}{4}\right) \chi + \left(1+\frac{\rho^2}{4}\right)\cos{t}} \label{conf0}\\
\tau &=& z \left(\frac{1}{\rho} + \frac{\rho}{4}\right)\sin{t}\\
R &=& z \left(\frac{1}{\rho}- \frac{\rho}{4}\right)\sqrt{1-\chi^2}\\
\Phi &=& \phi \label{conf3}
\eea
This maps the north pole of the $S^2$ to the origin of $R^2$, and at $t = 0$ the south pole is mapped to a point at infinity. At different $t$, the amount of the $S^2$ covered by the planar chart decreases, until at $t = \pm \pi$ only the north pole is covered.

We are here interested in the planar metric and stress tensor. Therefore it suffices to invert \eqref{conf0}-\eqref{conf3} in a near boundary expansion, developing expressions for $(t,\rho, \chi,\phi)$ in terms of $(\tau, R,\Phi)$ order-by-order in $z$. We find,
\bea
\rho &=& \frac{2}{V^{1/2}}z - \frac{2 (R^2-\tau^2)}{V^{3/2}}z^3 + \frac{2(R^4-\tau^2 + \tau^4 - R^2(1+2\tau^2))}{V^{5/2}}z^5+O(z^7)\nonumber \\
t &=& \arg\left(1+R^2-\tau^2+2i\tau\right) - \frac{2\tau}{V}z^2 + \frac{2\tau(1+R^2-\tau^2)}{V^2} z^4+O(z^6)\nonumber\\
\chi &=& \frac{1-R^2+\tau^2}{V^{1/2}} - \frac{4R^2z^2}{V^{3/2}} - \frac{6R^2(1-R^2+\tau^2)}{V^{5/2}}z^4+O(z^6)\nonumber\\
\phi & =& \Phi\label{inverted}
\eea
where we have defined $V\equiv (R^2 - \tau^2+ 1 )^2 + 4\tau^2$. Note that $V>0$.

\subsection{Applying to RT}
For axially-symmetric RT solutions the global metric in Fefferman-Graham gauge takes the form, 
\bea
ds_{(0)}^2 &=& -dt^2 + \frac{1}{\sigma(t,\chi)^2} d\Omega_2^2\\
ds_{(2)}^2 &=& g_{(2)ab}(t,\chi) dx^a dx^b\\
ds_{(3)}^2 &=& g_{(3)ab}(t,\chi) dx^a dx^b.
\eea
Applying the coordinate conversions \eqref{inverted} directly to the metric specified by these coefficients however leads to terms in the complete line element of the form, 
\be
a(\tau,R) dz^2, \qquad b(\tau,R)\frac{dz}{z} d\tau,  \qquad c(\tau,R)\frac{dz}{z}dR
\ee
i.e. outside the Fefferman-Graham gauge. The development of these terms can be compensated for order-by-order in $z$. Specifically, in addition to $\eqref{inverted}$ we apply the supplementary coordinate transformation, 
\bea
z &\to& z' = z\left(1+  z_2(\tau, R) z^2\right)\\
\tau &\to& \tau' = \tau + \tau_2(\tau,R)z^2\\
R &\to & R' = R + R_2(\tau,R)z^2
\eea
where, 
\bea
z_2(\tau, R) &=& \frac{R^2}{V} \left(\hat{\sigma}(\tau,R)^2-1\right)\\
\tau_2(\tau, R) &=& \frac{2R\tau}{V} \left(\hat{\sigma}(\tau,R)^2-1\right)\\
R_2(\tau, R) &=& \frac{R(1+R^2+\tau^2)}{V} \left(\hat{\sigma}(\tau,R)^2-1\right)\\
\hat{\sigma}(\tau,R) &\equiv& \sigma\left(\arg\left(1+R^2-\tau^2+2i\tau\right), \frac{1-R^2+\tau^2}{\sqrt{V}}\right),
\eea
then the cross terms vanish and we are in Fefferman-Graham gauge to high enough order in the $z$ expansion to extract the stress tensor. 
The boundary metric is given by, 
\bea
\tilde{g}_{(0)\tau\tau} &=& - \frac{1}{V}\left((1+R^2+\tau^2)^2 - 4 R^2\tau^2 \frac{1}{\hat{\sigma}(\tau,R)^2}\right)\\
\tilde{g}_{(0)\tau R} &=& - \frac{2 R \tau(1+R^2+\tau^2)}{V}\left( \frac{1}{\hat{\sigma}(\tau,R)^2}-1\right)\\
\tilde{g}_{(0)\tau \Phi} &=& 0\\
\tilde{g}_{(0)RR} &=&  \frac{R^2}{V}\left( \frac{(1+R^2+\tau^2)^2}{R^2}\frac{1}{\hat{\sigma}(\tau,R)^2} - 4 \tau^2\right)\\
\tilde{g}_{(0)R\Phi} &=& 0\\
\tilde{g}_{(0)\Phi\Phi} &=& \frac{R^2}{\hat{\sigma}(\tau,R)^2}
\eea
When $\sigma=1$ this reduces to Minkowski in polar coordinates.
Expanding near the origin of boundary polar coordinates ($R=0$) we see that the coordinate system is still regular.

The stress tensor is given in terms of the original metric components, $g_{(3)ab}(t,\chi)$. Again here we denote corresponding functions of the new leading order coordinates by $\hat{g}_{(3)ab}(\tau,R)$.
\bea
\tilde{g}_{(3)\tau\tau} &=& \frac{8(1+R^2+\tau^2)^2}{V^{5/2}} \hat{g}_{(3)tt}+ \frac{64 R^2 \tau(1+R^2+\tau^2)}{V^3} \hat{g}_{(3)t\chi}+ \frac{128 R^4 \tau^2}{V^{7/2}} \hat{g}_{(3)\chi\chi}\nonumber\\
\tilde{g}_{(3)\tau R} &=& -\frac{16 R\tau (1+R^2+\tau^2)}{V^{5/2}} \hat{g}_{(3)tt}\nonumber\\
&& - \frac{16 R(R^4+2R^2(1+3 \tau^2)+ (1+\tau^2)^2)}{V^3} \hat{g}_{(3)t\chi} - \frac{64 R^3 \tau (1+R^2 + \tau^2)}{V^{7/2}}\hat{g}_{(3)\chi\chi}\nonumber\\
\tilde{g}_{(3)\tau\Phi} &=& \frac{4(1+R^2+\tau^2)}{V^{3/2}} \hat{g}_{(3)t\phi} + \frac{16 R^2 \tau}{V^2} \hat{g}_{(3)\chi\phi}\nonumber\\
\tilde{g}_{(3)RR} &=& \frac{32 R^2 \tau^2}{V^{5/2}} \hat{g}_{(3)tt} + \frac{64 R^2 \tau (1+R^2+\tau^2)}{V^3} \hat{g}_{(3)t\chi} + \frac{32 R^2(1+R^2+\tau^2)^2}{V^{7/2}} \hat{g}_{(3)\chi\chi}\nonumber\\
\tilde{g}_{(3)R\Phi} &=& -\frac{8 R \tau}{V^{3/2}} \hat{g}_{(3)t\phi}- \frac{8 R (1+R^2+\tau^2)}{V^2} \hat{g}_{(3)\chi\phi}\nonumber\\
\tilde{g}_{(3)\Phi\Phi} &=& \frac{2}{V^{1/2}} \hat{g}_{(3)\phi\phi}\nonumber
\eea

The original stress tensor in the planar coordinate system is covariantly conserved and traceless, i.e.
\be
\nabla_{(0)}^a  g_{(3)ab} =0, \qquad g_{(0)}^{ab} g_{(3)ab} = 0 \label{originalConservation}
\ee
Here we verify that given the transformations above these conditions imply that the new planar-case stress tensor given by $\tilde{g}_{(3)}$ is covariantly conserved with metric $\tilde{g}_{(0)}$ and traceless. To use the relations derived from \eqref{originalConservation}, note that
\bea
\partial_\tau \hat{f}(\tau,R) &=& \frac{2(1+R^2+\tau^2)}{V}\partial_t f(t,\chi) + \frac{8 R^2 \tau}{V^{3/2}} \partial_\chi f(t,\chi) \bigg|_{(t(\tau,R),\chi(\tau,R))}\\
\partial_R \hat{f}(\tau,R) &=& -\frac{4R\tau}{V}\partial_t f(t,\chi) - \frac{ 4R (1+R^2+\tau^2)}{V^{3/2}} \partial_\chi f(t,\chi) \bigg|_{(t(\tau,R),\chi(\tau,R))}.
\eea
for some function $f$.
With these relations it is a straightforward to check that,
\be
\tilde{\nabla}_{(0)}^a  \tilde{g}_{(3)ab} =0, \qquad  \tilde{g}_{(0)}^{ab} \tilde{g}_{(3)ab} = 0 \label{newConservation}.
\ee

As a further technical note, in order to numerically compute the stress tensor in the planar case at some arbitrary value of $(R,\tau)$ we need to know the numerical solution, $\sigma(t,\chi)$ at,
\be
t = \arg\left(1+R^2-\tau^2+2i\tau\right)\qquad \chi =  \frac{1-R^2+\tau^2}{\sqrt{V}}.
\ee
Fortunately both are bounded, and we simply have to ensure that we have evaluated the solution in the domain $t\in [-\pi,\pi]$ and $\chi\in [-1,1]$. In practise we compute the solution numerically on a rectangular grid of  $(t,\chi)$-points covering this domain, then interpolate in order to extract the solution at the exact value of $(R,\tau)$ required.

\subsection{Examples}

\subsubsection{Schwarzschild black hole}
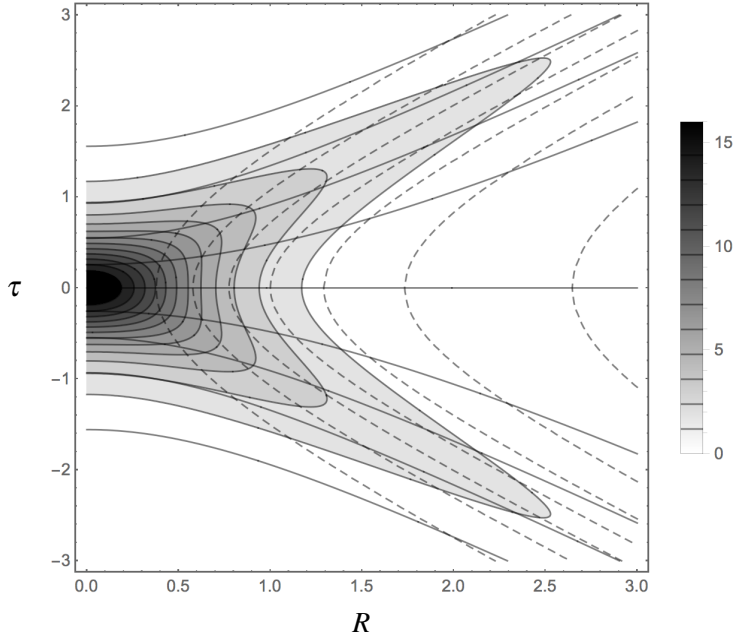
\begin{figure}[h!]
\begin{center}
\includegraphics[width=0.6\textwidth]{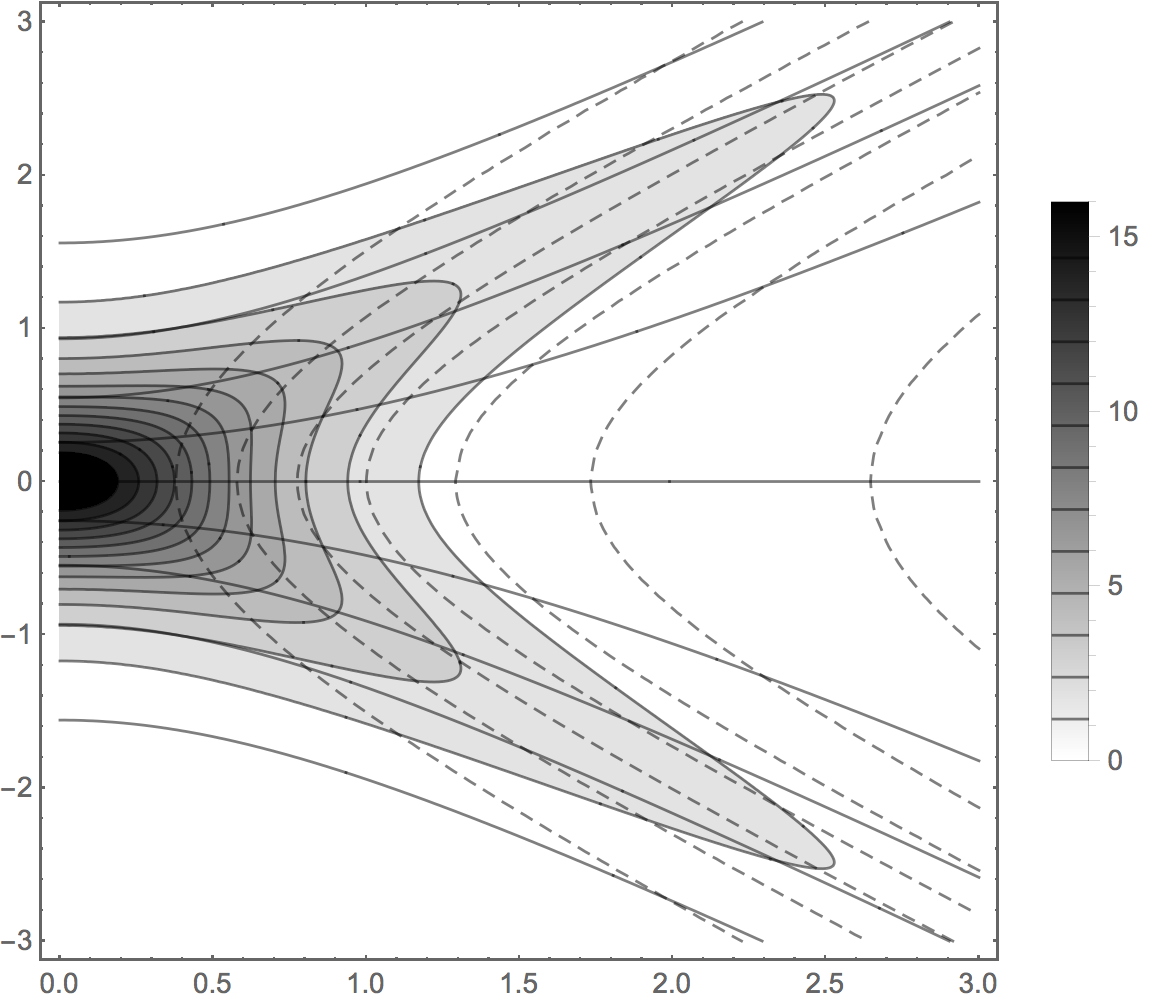}
\begin{picture}(0.1,0.1)(0,0)
\put(-145,-10){\makebox(0,0){$R$}}
\put(-275,116){\makebox(0,0){$\tau$}}
\end{picture}
\vspace{1em}
\caption{Contours of $\kappa^2\tilde{T}_{\tau\tau}/m$ for the global Schwarzschild (equilibrium) solution transformed to the Poincar\'e slicing; as in \cite{Friess:2006kw} but here we have one dimension fewer. The additional solid and dashed lines are the lines of constant $t$ and $\chi$ respectively in the global slicing.\label{csf}}
\end{center}
\end{figure}

Here we consider the simple case of the Schwarzschild solution corresponding to the trivial Calabi flow solution, $\sigma(t,\chi) = 1$, where we may work analytically. In global coordinates we have the boundary metric
\be
ds_{(0)}^2 = -dt^2 + \frac{d\chi^2}{1-\chi^2} + (1-\chi^2)d\phi^2
\ee
and stress tensor
\bea
\kappa^2 T_{tt} &=& 2m\\
\kappa^2 T_{\chi\chi} &=& \frac{m}{1-\chi^2}\\ 
\kappa^2 T_{\phi\phi} &=& m (1-\chi^2)
\eea
with other entries vanishing.
And so, in the planar coordinates we have the metric
\be
d\tilde{s}_{(0)}^2 = -d\tau^2 + dR^2 + R^2 d\Phi^2
\ee
and stress tensor
\bea
\kappa^2 \tilde{T}_{\tau\tau} &=& \frac{16 m(R^4+2R^2(1+2\tau^2) + (1+\tau^2)^2)}{V^{5/2}}\\
\kappa^2 \tilde{T}_{\tau R} &=& -\frac{48 m R \tau(1+R^2+\tau^2)}{V^{5/2}}\\
\kappa^2 \tilde{T}_{RR} &=& \frac{8 m(R^4+2R^2(1+5\tau^2) + (1+\tau^2)^2)}{V^{5/2}}\\
\kappa^2 \tilde{T}_{\Phi\Phi} &=& \frac{8 m R^2}{V^{3/2}}.
\eea
This situation is the 3d analogue of a flow considered in \cite{Friess:2006kw}. The evolution of the stress tensor is shown in figure \ref{csf}, illustrating how a shell of plasma collapses towards the origin when $\tau<0$, and then recedes for $\tau>0$ -- interpreted as the black hole passing through the Poincar\'e patch in the bulk.

\subsubsection{RT with an initial $\ell = 2$ inhomogeneity}
\begin{figure}[h!]
\begin{center}
\includegraphics[width=0.45\textwidth]{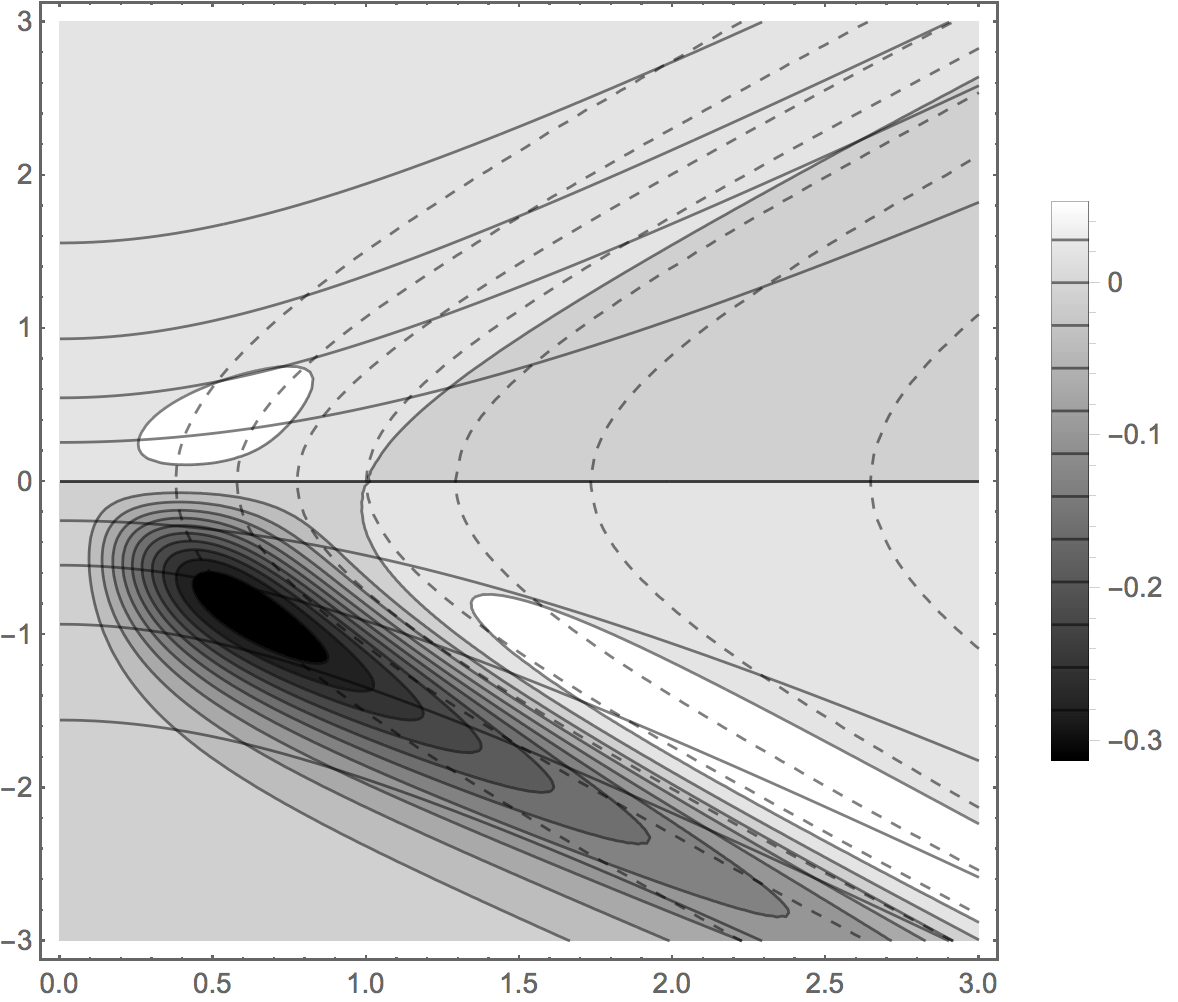}
\includegraphics[width=0.45\textwidth]{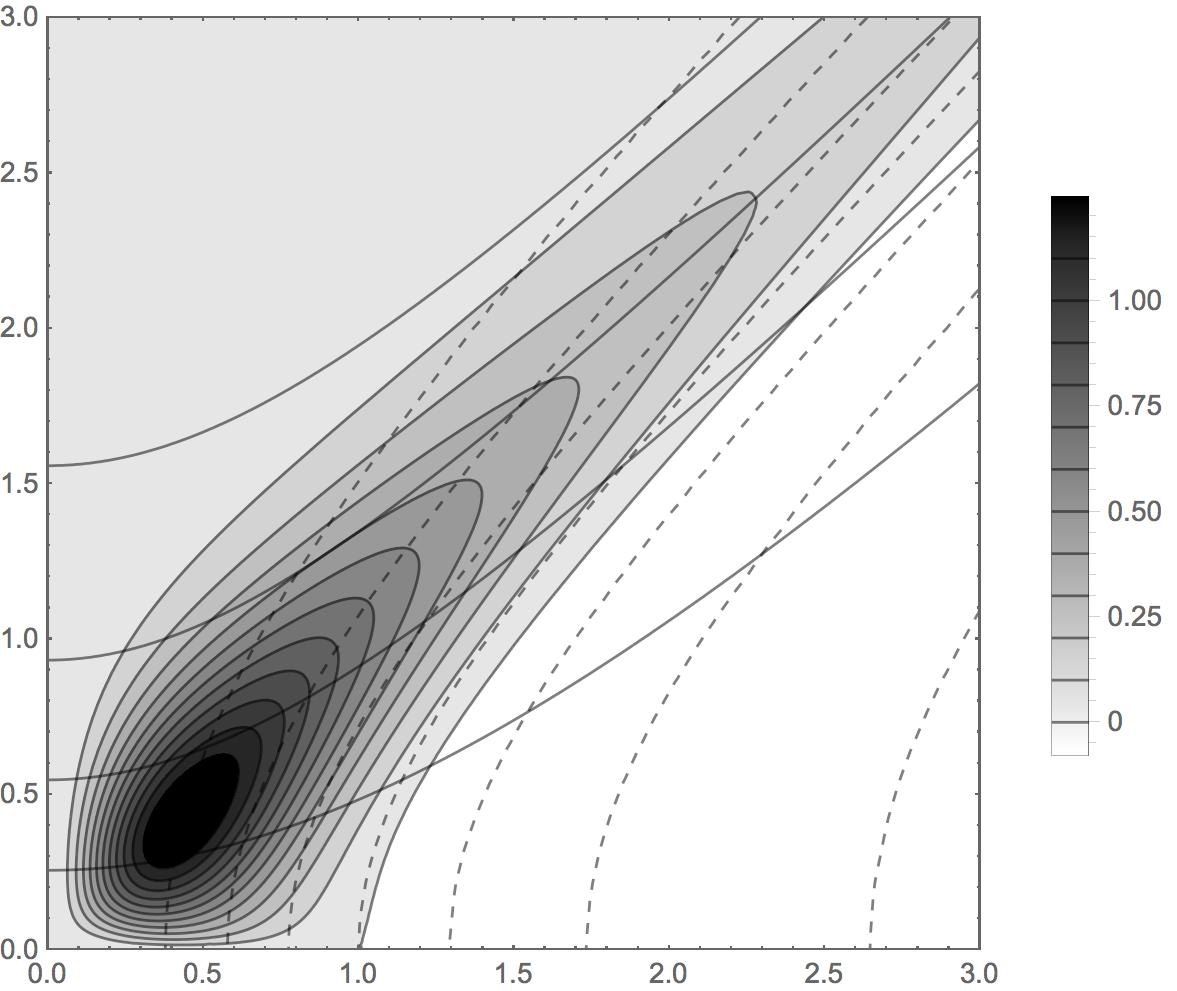}
\begin{picture}(0.1,0.1)(0,0)
\put(-112,-10){\makebox(0,0){$R$}}
\put(-308,-10){\makebox(0,0){$R$}}
\put(-404,85){\makebox(0,0){$\tau$}}
\end{picture}
\vspace{1em}
\caption{Deformations of the conformal plasma flow corresponding to an inhomogeneous RT spacetime. Shown are contours of deviations from Schwarzschild behaviour: $\kappa^2(\tilde{T}_{\tau\tau}-\tilde{T}^{\text{Schw.}}_{\tau\tau})/m$, obtained by reslicing a global RT solution in Poincar\'e patch-like coordinates as described in the text. The additional solid and dashed lines are the lines of constant $t$ and $\chi$ respectively in the global slicing. The RT spacetime evolves from an initial $\ell = 2$ inhomogeneity specified at $t=-\pi$ (left panel) and $t=0$ (right panel). \label{l2}}
\end{center}
\end{figure}
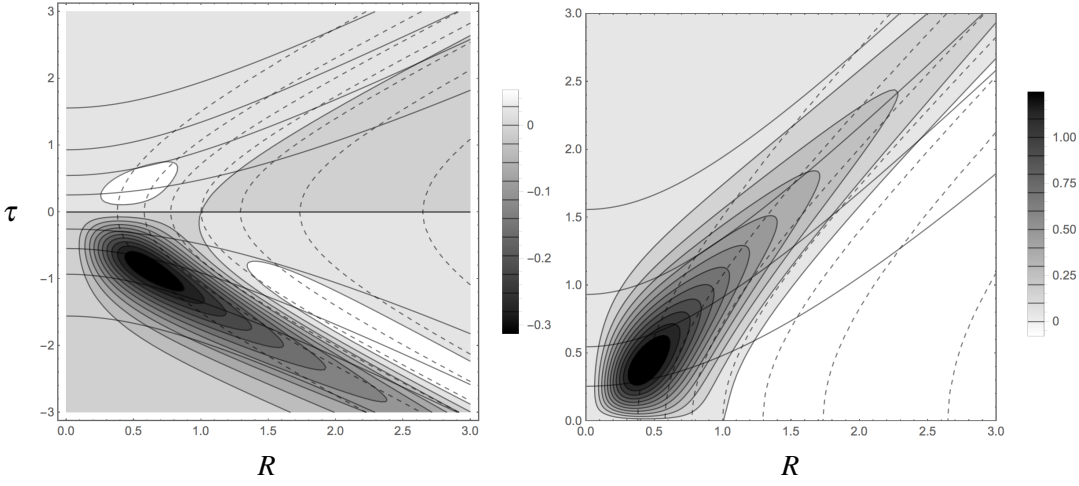

To illustrate the application to RT spacetimes we here consider the evolution of initial data of the form \eqref{initialdata} with $\ell = 2$. In linear theory the addition of an $\ell = 2$ perturbation to $\sigma$ corresponds to the addition of a quadrupole radiation component. We have freedom to choose the value of $t$ at which we begin the RT evolution, which through the coordinate transformation determines how much of the plasma flow on the plane we are describing. In the cases considered here we consider initial data at $t=-\pi$ and $t = 0$. A component of the holographic stress tensor on the plane, $\tilde{T}_{\tau\tau}$, is shown in figure \ref{l2} for $c = -0.8$.
 
\section{Discussion}
The RT spacetimes provide a convenient window into nonlinear, isometry-free solutions to the Einstein equations, including for $\Lambda < 0$. It is natural to try and utilise these solutions in the context of gauge/gravity duality, where they capture the dynamics of far-from-equilibrium plasmas in certain field theories. 

We first focussed on the applicability of the hydrodynamic approximation to the field theory duals to RT spacetimes, in the case where the conformal boundary is topologically $\mathbb{R}\times S^2$. The boundary metric and stress tensor can be extracted from the near boundary behaviour of the solutions. We proved that local fluid variables -- energy density and 4-velocity fields -- can be constructed to any order in a perturbative late time expansion. However, by analysing the non-perturbative regime using numerical solutions, we found that such variables cannot be identified everywhere for initial data which is sufficiently deformed from equilibrium. It would be interesting to try and resum the late time expansion to see these non-perturbative effects. At late times the solutions approach equilibrium at a rate governed by the longest lived mode identified in \cite{Bakas:2014kfa} corresponding to quadrupole radiation. We considered the nonlinear extension of this quadrupole mode to order 15 in a late time expansion, governed by a single amplitude $a$,  finding excellent agreement with the numerical data.

The regime described by linear quadrupole radiation is also a hydrodynamic regime \cite{Bakas:2014kfa}. The local fluid rest frame variables are well described by the linear quadrupole after a time, $t_{hyd}$, which we quantified. We found $t_{hyd} < t_{iso}$ where $t_{iso}$ is the time at which the transverse and longitudinal pressures become equal, consistent with other observations of hydrodynamisation in conformal theories \cite{Chesler:2010bi,Heller:2011ju}. We also constructed the quantity $t_{energy}$ describing deviation of the energy density from equilibrium (or here equivalently, the deviation of average pressure from equilibrium), and found that $t_{energy}<t_{hyd}$.

We also generated a new class of inhomogeneous plasma flows on the plane by performing coordinate transformations which correspond to moving to region analogous to the Poincar\'e patch in AdS. These coordinate transformations correspond to covering part of the boundary of topology $\mathbb{R}\times S^2$ with a chart of topology $\mathbb{R}^{1,2}$. We also showed how this transformation could be applied efficiently to solutions obtained numerically in global coordinates. We gave examples with $\ell=2$ initial inhomogeneities and the corresponding evolution of the expectation value of the dual stress tensor. Such solutions provide a way of introducing nonlinear deviations from the inhomogeneous flows provided by conformal transformations of the Schwarzschild solution, which do not themselves correspond to a RT solution on the plane.

\section*{Acknowledgements}
BW is supported by the NCCR under grant number 51NF40-141869 ``The Mathematics of Physics'' (SwissMAP). KS is supported in part by the Science and Technology Facilities Council (Consolidated Grant ``Exploring the Limits of the Standard Model and Beyond''). This project has received funding from the European Union's Horizon 2020 research and innovation programme under the Marie Skodowska-Curie grant agreement No 690575.
 
\bibliographystyle{JHEP}
\bibliography{proceedings} 
\end{document}